\documentclass[5p,twocolumn]{elsarticle}
\pdfoutput=1
\usepackage{amsmath,amssymb,amsbsy,multirow}
\usepackage[utf8]{inputenc}
\usepackage{graphicx}
\usepackage{ulem,color}
\usepackage[colorlinks=true, linkcolor=blue, filecolor=blue, urlcolor=blue, citecolor=blue, pdftex, plainpages=false]{hyperref}
\biboptions{sort&compress}

\newcommand{\be}{\ensuremath{\beta} }

\newcommand{\gc}{\ensuremath{g_c^2} }

\newcommand{\vev}[1]{\ensuremath{\left\langle #1 \right\rangle} }

\begin{document}
\title{Nonperturbative determination of $\beta$ functions for SU(3) gauge theories \newline with 10 and 12 fundamental flavors using domain wall fermions}

\author[CO]{A.~Hasenfratz\corref{cor1}}
\ead{anna.hasenfratz@colorado.edu}
\author[BU]{C.~Rebbi}
\ead{rebbi@bu.edu}
\author[UoE,CO]{O.~Witzel}
\ead{oliver.witzel@colorado.edu}
\cortext[cor1]{Corresponding author}
\address[CO]{Department of Physics, University of Colorado, Boulder, CO 80309, USA}
\address[BU]{Department of Physics and Center for Computational Science, Boston University, Boston, MA 02215, USA}
\address[UoE]{Higgs Centre for Theoretical Physics, School of Physics \& Astronomy, The University of Edinburgh, Edinburgh, EH9 3FD, UK}

\date{\today}
\begin{abstract}%
  Nonperturbative lattice field theory simulations provide a systematic framework to investigate properties of conformal systems at strong couplings. These simulations  can be performed using different lattice  discretizations.  Here we present numerical results for the step scaling beta function in SU(3) gauge theories with ten and twelve fundamental flavors. We calculate the renormalized $\beta$  function in the finite volume gradient flow renormalization scheme. Using M\"obius domain wall fermions with Symanzik gauge action,  Zeuthen gradient flow, and perturbative tree-level improvement, we implement a fully $O(a^2)$ Symanzik improved set-up and demonstrate its advantages. We compare our findings to existing results in the literature.  For the ten flavor system we observe excellent agreement with the domain wall step-scaling function calculated by Chiu for the range in $g_c^2$ where our data overlap.  In the case of the twelve flavor system, our $O(a^2)$ Symanzik improved set-up predicts a conformal infrared fixed point around $g_c^2 \sim 5.5$ in the $c=0.25$ scheme, which is presently in tension with staggered fermion results in the literature.  We consider possible reasons for the discrepancy. 
\end{abstract}

\maketitle
\section{Introduction}\label{intro}

The Higgs sector and electroweak symmetry breaking in the Standard Model could  be described by models based on a strongly coupled gauge theory which is close to a conformal fixed point (FP) \cite{Luty:2004ye,Dietrich:2006cm,Vecchi:2015fma,Ferretti:2013kya,Brower:2015owo,Hasenfratz:2016gut}. Such a gauge theory is governed by an infrared fixed-point (IRFP) where the gauge coupling becomes irrelevant and the mass is the only relevant operator. Conformal systems  allow to define two different continuum limits on the $m=0$ critical surface: 1) If the gauge coupling is tuned to zero, the continuum limit is governed by the Gaussian fixed point (GFP). The renormalized coupling can be kept fixed at any finite value up to the IRFP. This is very similar to the 3-dimensional scalar model when the quartic coupling is tuned to zero, describing a tricritical system.   2) Alternatively, it is possible to define a continuum limit by not tuning the gauge coupling. In the IR limit, the gauge coupling runs to the IRFP, its renormalized value is finite and independent of the bare coupling at the cut-off.  This is similar  to the 3-dimensional scalar model when only the mass is tuned to the critical  surface  and the quartic coupling is allowed to run to the Wilson-Fisher fixed point (WFFP).

Establishing the presence of an IRFP in 4-dimensional gauge-fermion systems is a challenging problem. Perturbative calculations are justified near the GFP at $g^2=0$ but become unreliable at stronger couplings where an IRFP might occur. Numerical lattice field theory simulations allow to explore  the strong coupling domain. The basis of lattice investigations is the concept of universality which ensures that in the infinite cut-off ``continuum'' limit the details of the lattice discretization (action) are irrelevant, as long as the continuum symmetries of the system are preserved or restored. Universality is well understood in asymptotically free QCD-like systems where the continuum limit is defined around  $g^2=0$ \cite{Golterman:1984cy,Sharpe:2006re}. We also expect universality to be preserved along the renormalized trajectory (RT) emerging from the GFP to $g^2>0$ which corresponds to a ``perfect'' lattice action without any artifacts \cite{Hasenfratz:1993sp,DeGrand:1995ji,Bietenholz:1996qc}. The conformal IRFP sits on the RT and therefore also exhibits correct continuum symmetries without lattice artifacts. This, however, does not exclude the possibility of other FPs at non-vanishing gauge couplings. Therefore,  the continuum limit of numerical simulations has to be taken with care to ensure it is in the basin of attraction of the FPs and thus approaches the RT. By numerically determining the renormalization group (RG) $\beta$ function, we aim to explore the nature and, if possible, the existence of an IRFP for SU(3) gauge systems with ten and twelve flavors.

The  RG $\beta$-function  depends on the details of the regularization and renormalization prescription. By fixing the renormalization scheme,  the $\beta$-function of a renormalized coupling becomes independent of the regularization. For example, the finite volume gradient flow (GF) step scaling function, the lattice  analogue of the continuum RG $\beta$-function, depends only on the renormalization scheme, but not on the lattice discretization.

In the case of the SU(3) gauge system with twelve fundamental flavors, several groups investigated the GF step scaling function using different discretizations of staggered fermions but arrived at different conclusion about the IR properties \cite{Lin:2015zpa,Hasenfratz:2016dou,Fodor:2016zil,Fodor:2017gtj}.
To understand these discrepancies, we calculate  the  finite volume GF step  scaling function  of the 10 and 12 flavor SU(3) gauge systems with domain wall (DW) fermions. Using the fully $O(a^2)$ Symanzik improved combination of tree-level improved Symanzik gauge action,  Zeuthen gradient flow,  and tree-level improved Symanzik operator \cite{Ramos:2014kka, Ramos:2015baa}, we significantly reduce discretization effects.  The effectiveness of the perturbative improvements imply that our simulations are close to the GFP, supporting the reliability of our continuum limit extrapolations.  These extrapolations are performed with up to five lattice volume pairs and lattice volumes up to $32^4$.

The main objective of this paper is to study  the $N_f=12$ flavor system with domain wall fermions. In the $c=0.25$ renormalization scheme, our preferred set-up predicts the step scaling function up to $g_c^2\lesssim 6.25$,  and an IRFP at $g_c^2 \sim 5.5$.  Since this is the first DW calculation of this system, we have complemented our study by investigating a system with 10  flavors in a limited coupling range. Our 10 flavor result  agrees well with the DW  prediction of Refs.~\cite{Chiu:2016uui,Chiu:2017kza,Chiu:2018edw},  boosting confidence in our results.

The remainder of this paper is organized as follows: In Sec.~\ref{sec:GF} we define the gradient flow discrete $\beta$ function and how it is calculated on the lattice. We provide details of our numerical simulations in Sec.~\ref{sec:NumSim} and present our analysis in Sec.~\ref{sec:StepScaling}. First we demonstrate our analysis and present results for the twelve flavor discrete $\beta$-function  before showing results for $N_f=10$. Subsequently we discuss our findings in Sec.~\ref{sec:Discussion} and compare them to non-perturbative and perturbative results published in the literature.
 Preliminary results of this work were presented at Lattice 2017, 2018, and 2019 \cite{Hasenfratz:2017mdh,Hasenfratz:2018wpq,Nf12stepScaling}.

\section{\texorpdfstring{Gradient flow discrete $\be$ function}{Gradient flow discrete beta function}} \label{sec:GF}

The continuum step scaling function, or discrete $\beta$ function is the infinite volume $(a / L) \to 0$  limit of the finite volume discrete $\be$ function corresponding to  a  change of scale by a factor  $s$ 
\begin{equation}
  \label{eq:beta}
  \be_{c,s}(\gc; L) = \frac{\gc(sL; a) - \gc(L; a)}{\log(s^2)},
\end{equation}
where $\gc(L;a)$ is a  renormalized coupling at the energy scale set by the volume $\mu = (c L )^{-1}$, and $c$ is a parameter that defines the renormalization scheme. Gradient flow  can be used  to define $\gc(L)$ \cite{Narayanan:2006rf,Luscher:2010iy,Fodor:2012td, Fodor:2012qh, Fritzsch:2013je} as
\begin{equation}
  \label{eq:pert_g2}
  \gc(L) = \frac{128\pi^2}{3(N^2 - 1)} \frac{1}{C(c,L)} \vev{t^2 E(t)}
\end{equation}
where  the flow time $t$ is related to the energy scale as $\mu = 1/ \sqrt{8 t}$, $E(t)$ is the energy density at $t$, and $C(c,L)$  denotes a normalization factor that removes tree-level discretization effects \cite{Fodor:2014cpa}.  The continuum limit extrapolated discrete $\beta$-function $\beta_{c,s}(g^2_c) = \lim_{(a / L) \to 0} \beta_{c,s}(g^2_c, L)$ depends only on the renormalized coupling $g^2_c$. 
Therefore it  is expected to be  independent of  irrelevant operators introduced by the lattice regularization and depends only on the renormalization parameters such as $c$ and $s$. 

Cut-off effects on $\gc(L;a)$ and $ \be_{c,s}(\gc; L)$  originate from the discretization of the lattice action, the gradient flow transformation, and the operator used for measuring $\vev{E(t)}$. Full $O(a^2)$ Symanzik improvement is possible  by using the L{\"u}scher-Weisz tree-level improved gauge action, also referred to as Symanzik  action, the Zeuthen gradient flow, and the  L{\"u}scher-Weisz (or Symanzik) operator to measure the energy density \cite{Ramos:2014kka, Ramos:2015baa}.

In our simulations we  generated gauge configurations at the same set of bare gauge couplings $\{\be\}$ on all volumes. Hence we can calculate  $\be_{c,s}(\gc;L)$ directly at every $\be$. Since we take the difference of two independently  predicted  $\gc(sL;a)$ and $\gc(L;a)$  values, the errors can be reliably estimated. We interpolate the resulting finite volume step scaling function  $\be_{c,s}(\gc;L)$ polynomially as
\begin{equation}
\be_{c,s}(\gc;L) = \sum_{i=0}^{n} b_i g_c^{2i},
\label{eq:fit_form}
\end{equation}
which is motivated by the perturbative expansion. We find  that $n=3$  provides a good interpolation form. When the finite volume step scaling functions have small cut-off effects, as is the case with Zeuthen flow and  tree-level normalized Symanzik operator,  we find that $b_0 =0 $ within errors and hence implement that as constraint. 
The continuum $(a/L)^2\to 0$ extrapolation is performed on these fitted functional forms.

\section{Numerical simulations}\label{sec:NumSim}
To determine the $\beta$-function nonperturbatively, we generate ensembles of gauge field configurations with either ten or twelve dynamical flavors using M{\"o}bius domain wall fermions \cite{Brower:2012vk} with three levels of stout smearing \cite{Morningstar:2003gk} in combination with tree-level Symanzik (L\"uscher-Weisz (LW)) gauge action \cite{Luscher:1984xn,Luscher:1985zq}.\footnote{Properties of this combination of gauge and fermion actions have been explored for QCD simulations in Refs.~\cite{Kaneko:2013jla,Noaki:2015xpx}.} The gauge fields are generated using the \texttt{Grid} code \cite{Boyle:2015tjk,GRID} with trajectory length $\tau=2$ Molecular Dynamics  Time Units (MDTU) and we choose periodic boundary conditions (BC) for the gauge field and anti-periodic BC for the fermion fields in all four space-time directions.

Chiral DW-fermions are simulated with an additional fifth dimension of extent $L_s$ separating the physical modes of the four dimensional space-time. In practice $L_s$ is necessarily finite which leads to residual chiral symmetry breaking parametrized by an additive mass $a m_\text{res}$. For our simulations we set the bare fermion mass to zero and monitor the residual chiral symmetry breaking by determining $m_\text{res}$ numerically from the midpoint-pseudoscalar correlator. By increasing $L_s$ at the strongest couplings, we ensure that $am_\text{res}$ is sufficiently small, $5\cdot 10^{-6}$ or less, to be negligible for our analysis. Details are reported in \ref{Appendix.mres}.

Our ensembles are symmetric $L^4$ volumes with $L/a$ ranging from 8 to 32.   For each ensemble  we typically collected 3-5k  MDTU for $N_f=10$ and 7-10k MDTU for $N_f=12$, but only 2-4k MDTU on the largest $28^4$ and $32^4$ volumes.  We identified a first order bulk phase transition for both  $N_f=10$ and 12 around $\beta=4.00$. When we  approach this phase transition, the residual mass increases rapidly, limiting the range of couplings we can simulate.   

Measurements of the gradient flow are performed using \texttt{qlua} \cite{Pochinsky:2008zz,qlua} on configurations separated by 10 MDTU.  In addition to Zeuthen flow \cite{Ramos:2014kka, Ramos:2015baa} we also compute Wilson (W) and Symanzik (S) flows to estimate systematic effects. In all cases we estimate the energy density by measuring the Wilson-plaquette (W), clover (C), and  Symanzik (S) operators. Our data are analyzed using the $\Gamma$-method \cite{Wolff:2003sm} to estimate  and account for autocorrelation times which mostly are around 20 MDTU but always less than 100 MDTU.

\section{Analysis of the step scaling function}\label{sec:StepScaling}
As mentioned in Section \ref{sec:GF}, our preferred analysis is based on  the  Zeuthen flow combined with the  S operator. Combined with the Symanzik gauge action of our simulation, this choice is fully ${\cal O }(a^2)$ Symanzik improved \cite{Ramos:2014kka}. Further we include the tree-level normalization factor in the gradient flow coupling to reduce remaining cut-off effects.  We refer to this analysis as ``nZS'' and compare  our results to  alternative determinations.\footnote{Our shorthand for the different analyses denotes with a capital letter the gradient flow (S,W,Z) followed by the operator (C,S,W). When using tree-level normalization, we prefix a lower case 'n'.} For $N_f=12$ we compare our results to  determinations based on Symanzik flow with clover operator (without tree-level normalization), abbreviated by ``SC'' and the choice in Ref.~\cite{Fodor:2017gtj}, tree-level normalized Wilson flow with clover operator ``nWC", used in Ref.~\cite{Hasenfratz:2016dou} together with the Wilson gauge action, and also to ``ZS" without tree-level normalization. For $N_f=10$ we base the alternative determination on ``nWC'' to match the choice in Refs.~\cite{Chiu:2016uui,Chiu:2017kza,Chiu:2018edw} which however uses a different gauge action.

With respect to our preliminary results \cite{Hasenfratz:2017mdh}, we have extended our data set to include larger volumes and, for $N_f=12,$ have increased statistics and performed simulations at stronger couplings.  In addition, the use of tree-level normalization considerably reduces discretization artifacts. Preliminary results were presented in Refs.~\cite{Hasenfratz:2017mdh,Hasenfratz:2018wpq,Nf12stepScaling}.

\subsection{\texorpdfstring{Step-scaling function for $N_f=12$}{Step-scaling function for Nf=12 }}

\begin{figure}[tbp] 
  \centering
  \includegraphics[width=0.95\columnwidth]{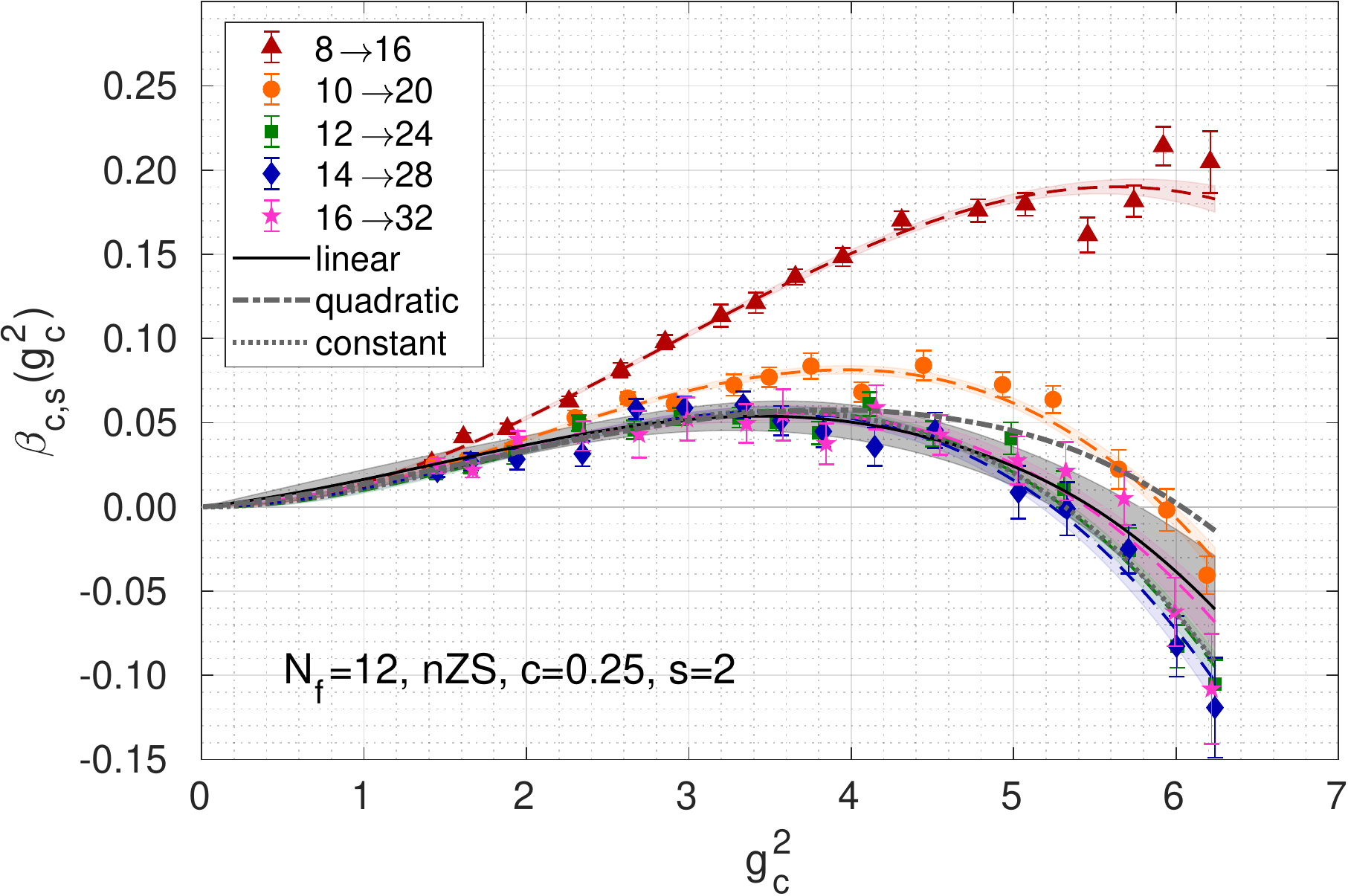}\\
  \includegraphics[width=0.95\columnwidth]{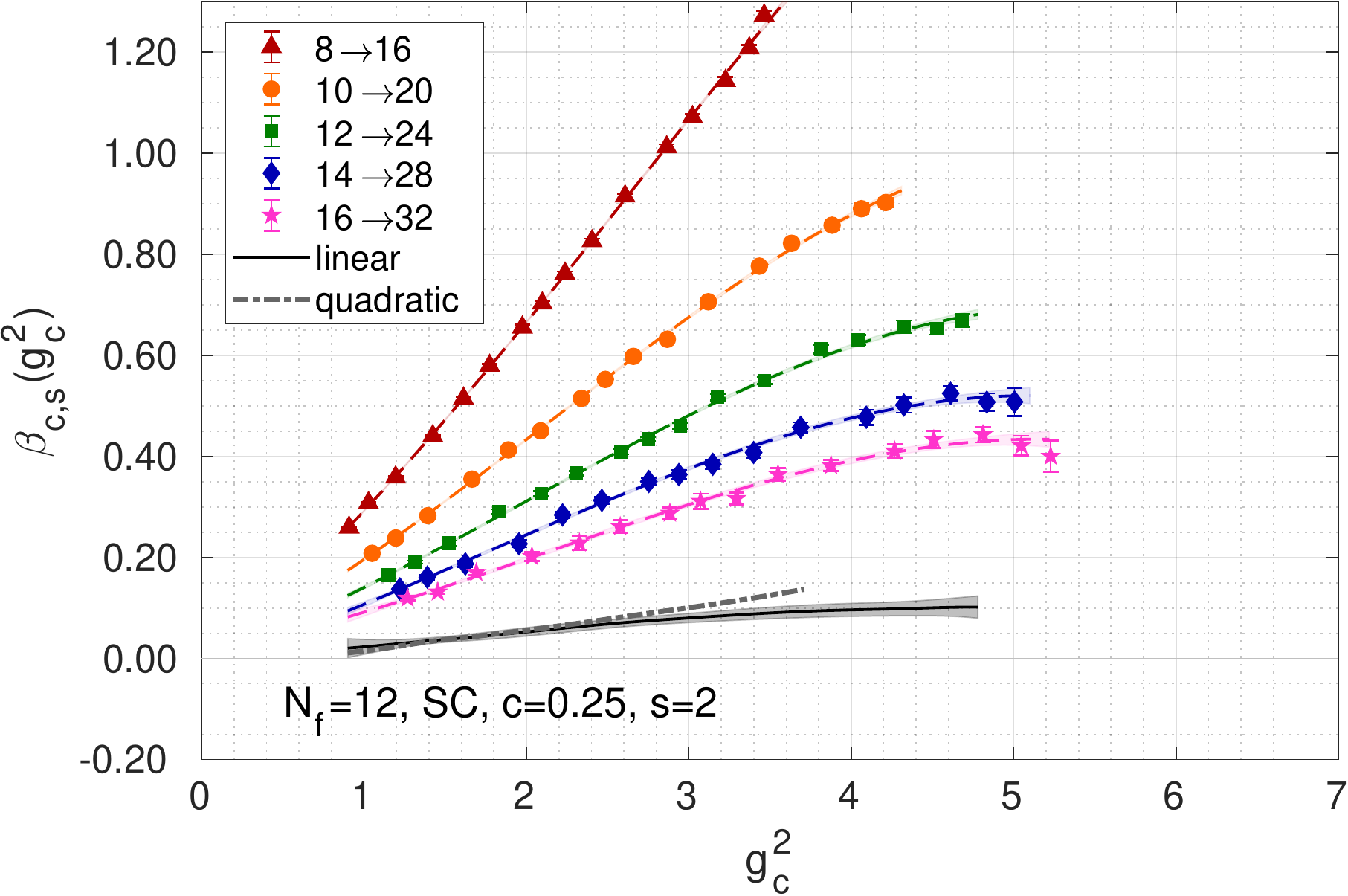}\\  
  \includegraphics[width=0.95\columnwidth]{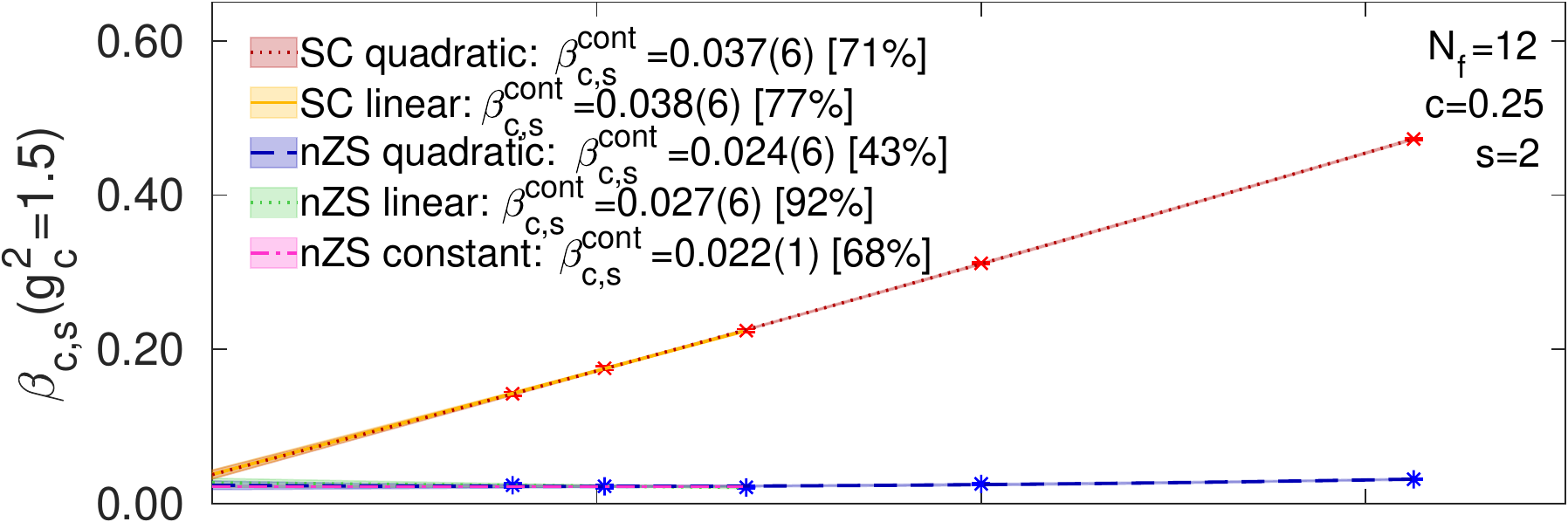}\\
  \includegraphics[width=0.95\columnwidth]{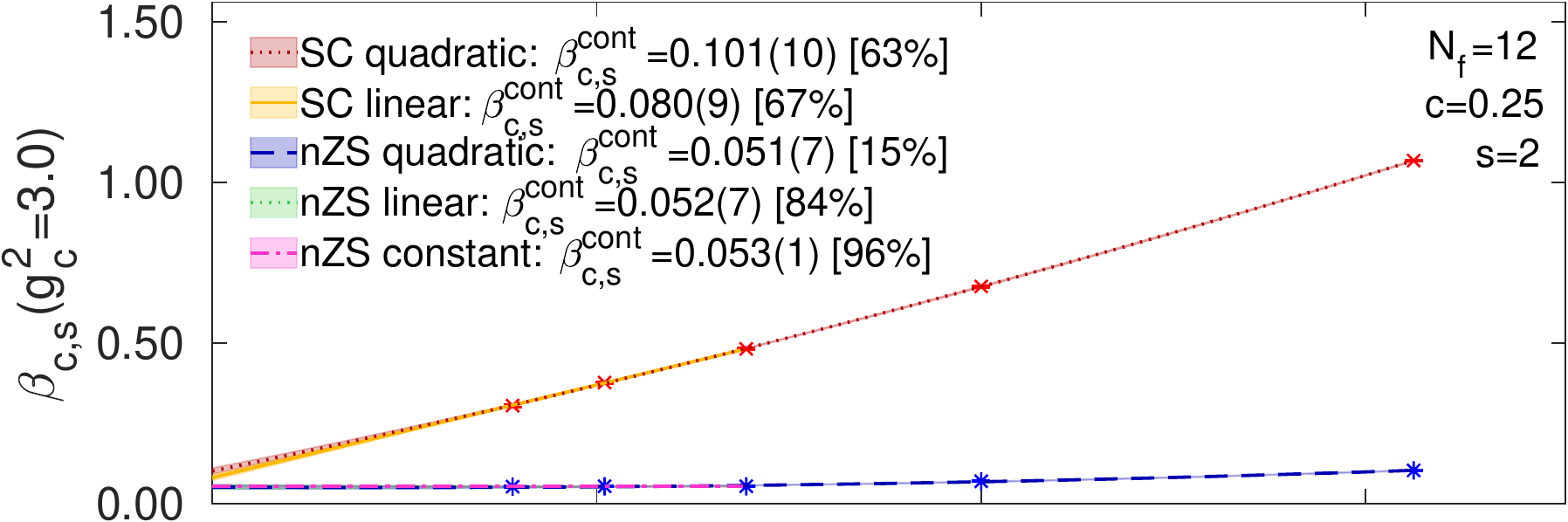}\\
  \includegraphics[width=0.95\columnwidth]{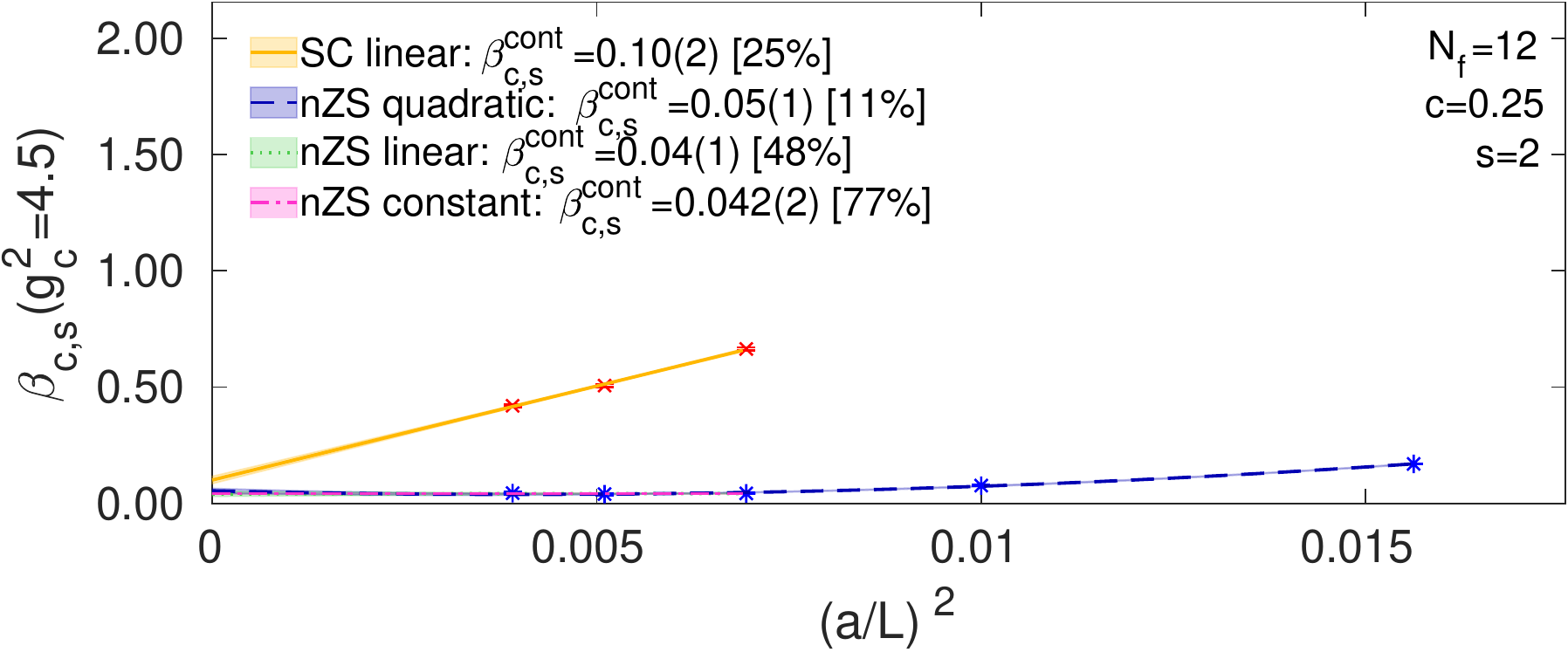}
  \caption{Step scaling functions for degenerate 12 flavors analyzed  in the renormalization scheme $c=0.25$. Colored symbols and bands correspond to individual lattice volume-pairs; the black line with gray band shows the continuum-limit discrete $\beta$-functions obtained from a linear extrapolation of the three larges lattice volume pairs. Alternative extrapolations are indicated by gray dash-dotted or dotted lines. The top  two panels show nZS and SC data, while details of the continuum extrapolations are presented in the three lower panels for $g_c^2=1.5$, 3.0, 4.5 with $p$-values in percent given in brackets. (Error bars and bands reflect only statistical uncertainties.)}
  \label{fig:Nf12}
\end{figure}

Motivated by the published results based on staggered fermions \cite{Hasenfratz:2016dou,Fodor:2017gtj}, we  use $c=0.25$ in our analysis. Choosing the scale factor $s=2$, we pair the lattice volumes $8\to 16$, $10\to 20$, $12\to 24$, $14\to28$, and $16\to 32$ to calculate the discrete $\beta_{c,s}$-function according to Eq.~(\ref{eq:beta}). Based on $g_c^2$ determined from Zeuthen flow, Symanzik operator, and tree-level improvement (nZS), we obtain the five sets of colored data points in the top panel of Fig.~\ref{fig:Nf12}.  Using a third order polynomial in $g_c^2$, we interpolate these data points and observe that the discretization effects are too small to resolve the intercept at $g^2_c=0$. Hence we constrain the intercept to be zero and show the resulting, interpolating curves by the dashed lines with shaded error bands in the same color as the data points. The quality of the interpolating fits is reflected by these bands passing through almost all data points.  In the weak coupling limit the perturbative improvement works perfectly, the data points sit on top of each other, i.e.~$(a/L)^2$, and even most of the higher order effects, are removed. For stronger couplings, the improvement still reduces artifacts, but in particular the small lattice volumes exhibit increasing discretization effects.  In a final step, we obtain the continuum limit $\beta_{c,s}$-function  by extrapolating these interpolated step-scaling functions in $(a / L)^2$  at fixed $g_c^2$. We consider  three different continuum limit extrapolations: quadratic ansatz in $(a/L)^2$ to fit all five lattice volume pairs (dashed-dotted gray line), linear fit in $(a/L)^2$ using only the three largest lattice volume pairs (solid black line with gray error band), and a constant fit to the three largest lattice volume pairs (gray dotted line). All three extrapolations are consistent at the $1\sigma$ level and exhibit good $p$-values, mostly far greater than $10\%$, throughout the $g_c^2$ range covered.

The second panel of Fig.~\ref{fig:Nf12}  shows the result of the same analysis with Symanzik flow and clover operator without tree-level normalization (SC). The SC combination covers a  smaller range in $g_c^2$ and has larger discretization effects. We therefore consider only a quadratic interpolation in $(a/L)^2$ for $g_c^2<3.6$ and a linear extrapolation for $g_c^2\lesssim 4.8$.

For selected values of $g_c^2$, we present details on the continuum extrapolation in the lower three panels of Fig.~\ref{fig:Nf12} where we compare up to five different extrapolations: constant, linear, and quadratic extrapolations of our preferred nZS data are contrasted with the alternative SC combination. Since the SC data covers a narrower $g_c^2$ range, the quadratic extrapolation of SC is missing in the last panel. The comparison reveals that our nZS data exhibit extremely small discretization effects which result in a very flat continuum extrapolation. In fact, extrapolating the three largest lattice volumes by just fitting a constant has excellent $p$-values. Since nZS is fully ${\cal O}(a^2)$ improved, the small cut-off effects also imply that the simulations are performed close to the perturbative regime.

In contrast to nZS, SC data exhibit significantly larger discretization effects which  grow substantially for stronger couplings. These discretization effects pull the different lattice volume pairs further apart and force the continuum extrapolation to cover a much larger range. While for weak couplings, linear and quadratic continuum extrapolations of SC data are consistent, they differ for $g_c^2\gtrsim 3.0$. This indicates that in case of the SC analysis, the present data set is insufficient to take a reliable continuum limit which also explains why the continuum limit of nZS and SC data differs beyond the $2\sigma$ level for stronger couplings. The SC data demonstrate that discretization effects of the GF coupling can differ significantly on the same set of ensembles.   For our set of ensembles, the shown SC data are a worst-case example for the total of 18 different determinations\footnote{As mentioned before we have determined renormalized coupling for Zeuthen, Symanzik, and Wilson flow using the Wilson-plaquette, clover, and Symanzik operator with or without tree-level normalization.}  we consider. Linear and quadratic continuum extrapolations of our SC data  are inconsistent for $g_c^2\gtrsim 3.0$, rendering the SC extrapolations to be  untrustworthy. Reducing discretization effects by applying tree-level improvement (i.e.~nSC analysis),  the predicted $\beta$ function approaches our preferred nZS determination and becomes consistent  at the $ \sim 1.5\sigma$ level~\cite{Nf12stepScaling}.

 In addition we consider nWC and observe discretization effects which are of similar size as for nZS. The predicted nWC continuum limit is consistent at the $\sim 1\sigma$ level with our nZS determination. Corresponding plots for the nWC analysis are shown in Fig.~\ref{fig:Nf12_nWC} in \ref{Sec.Nf12nWC}.

\subsection{\texorpdfstring{Step-scaling function for $N_f=10$}{Step-scaling function for Nf=10}}

\begin{figure}[tbp] 
  \centering
  \includegraphics[width=0.95\columnwidth]{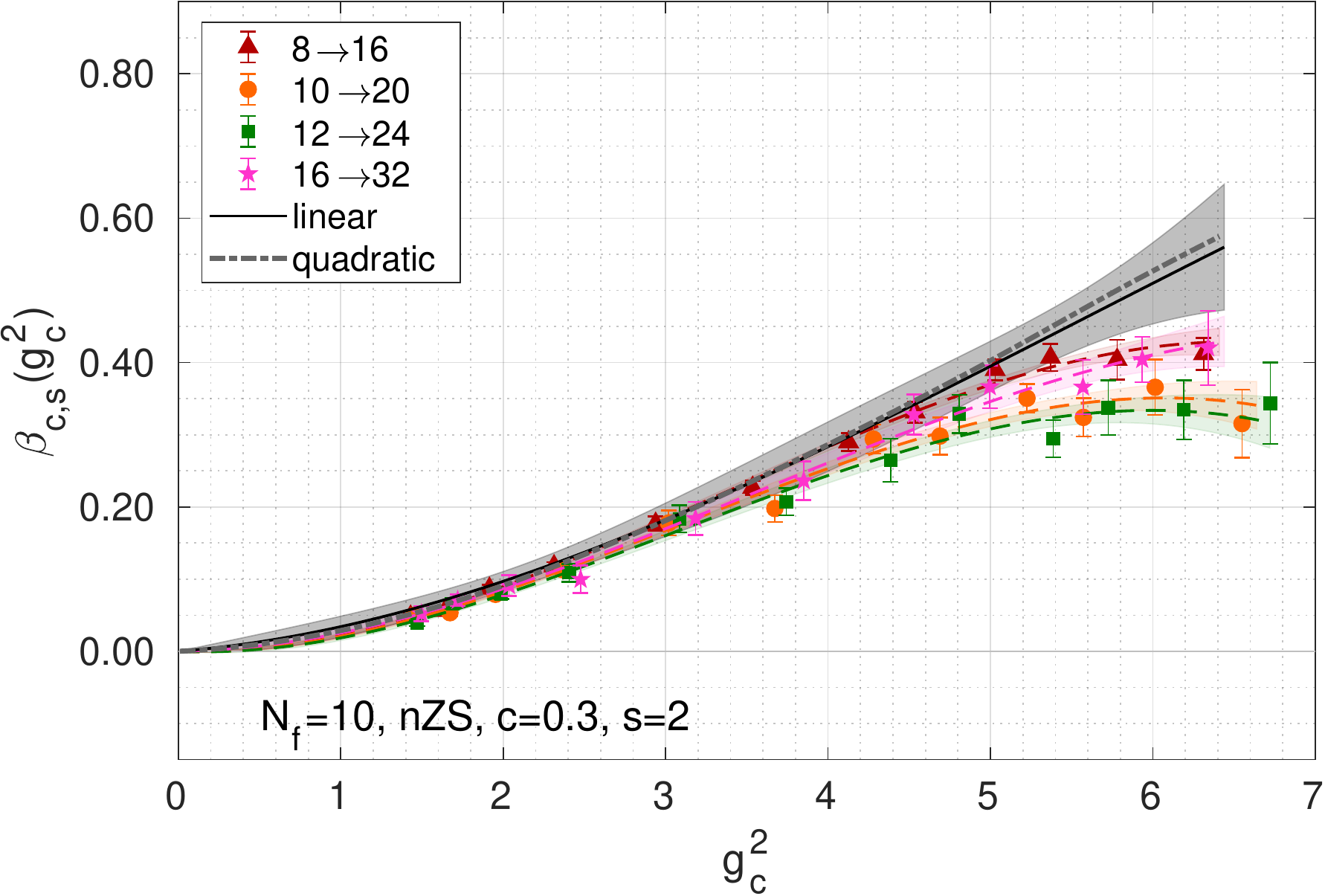}
  \includegraphics[width=0.95\columnwidth]{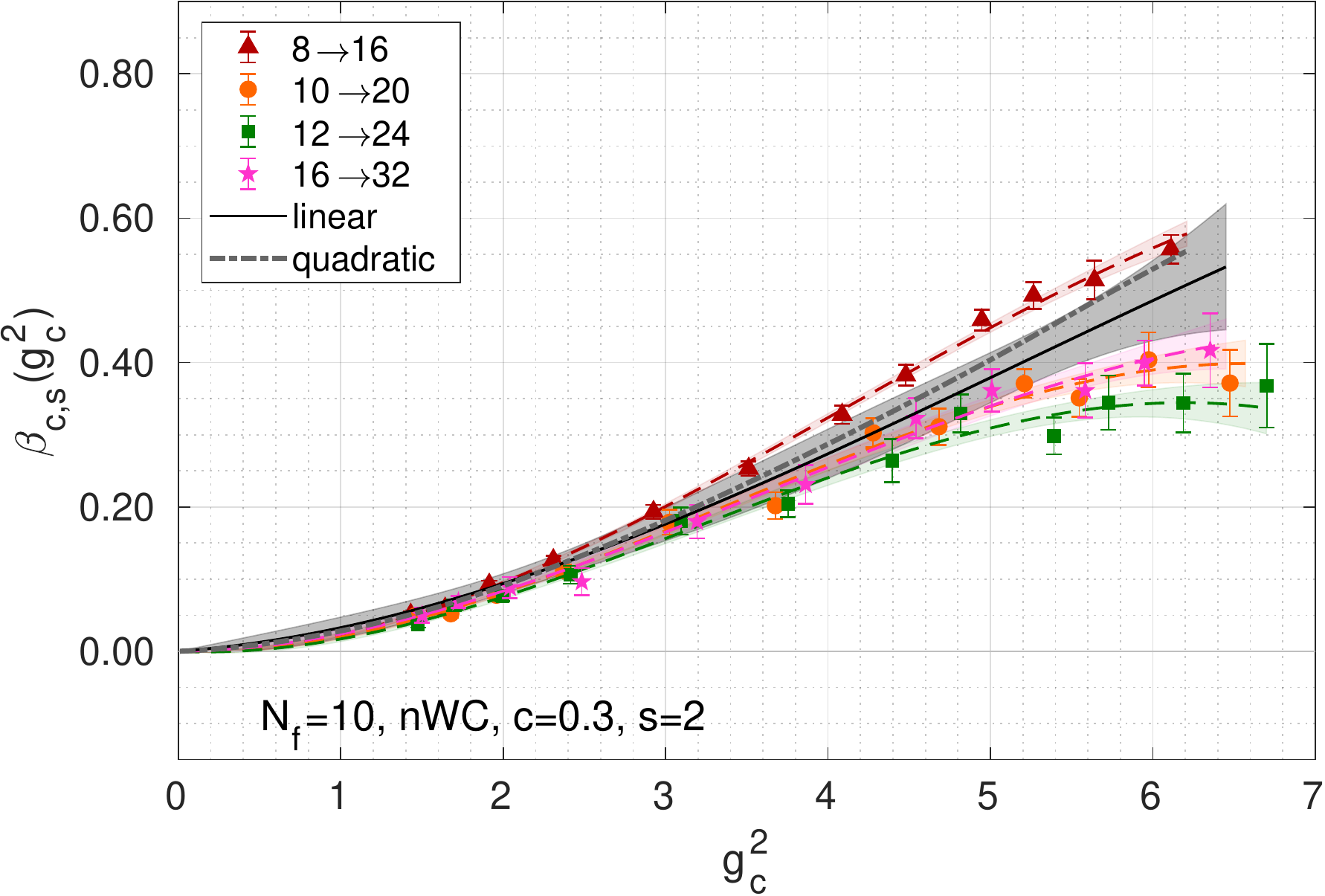}\\  
  \includegraphics[width=0.95\columnwidth]{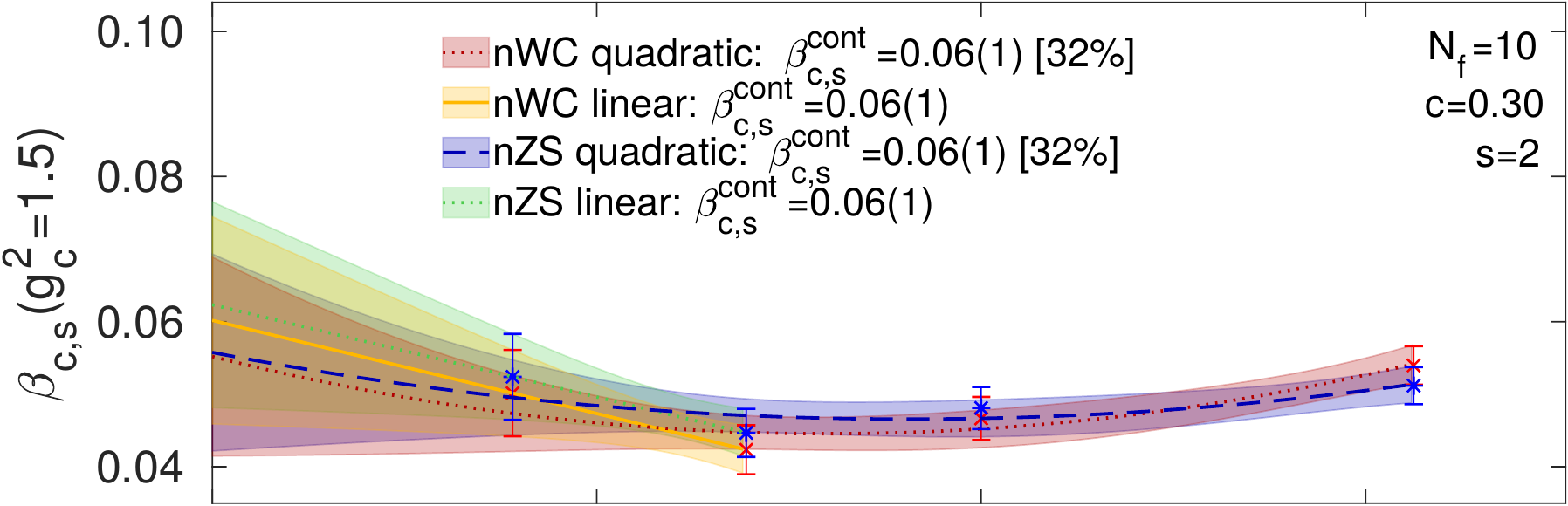}\\
  \includegraphics[width=0.95\columnwidth]{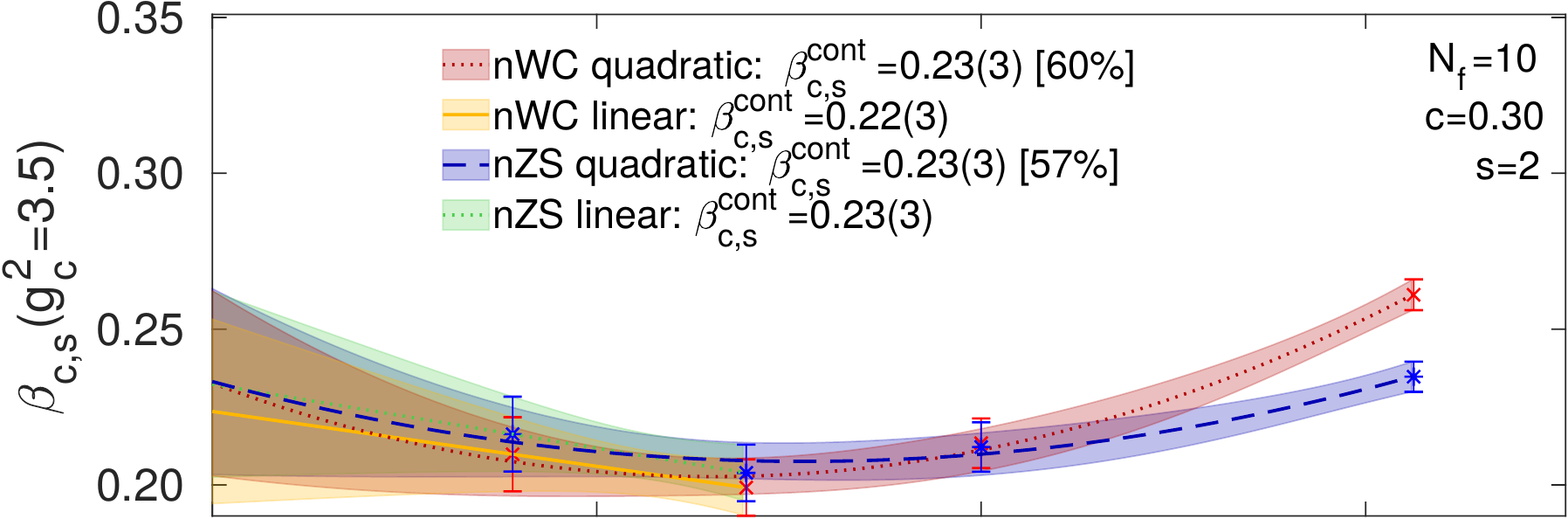}\\
  \includegraphics[width=0.95\columnwidth]{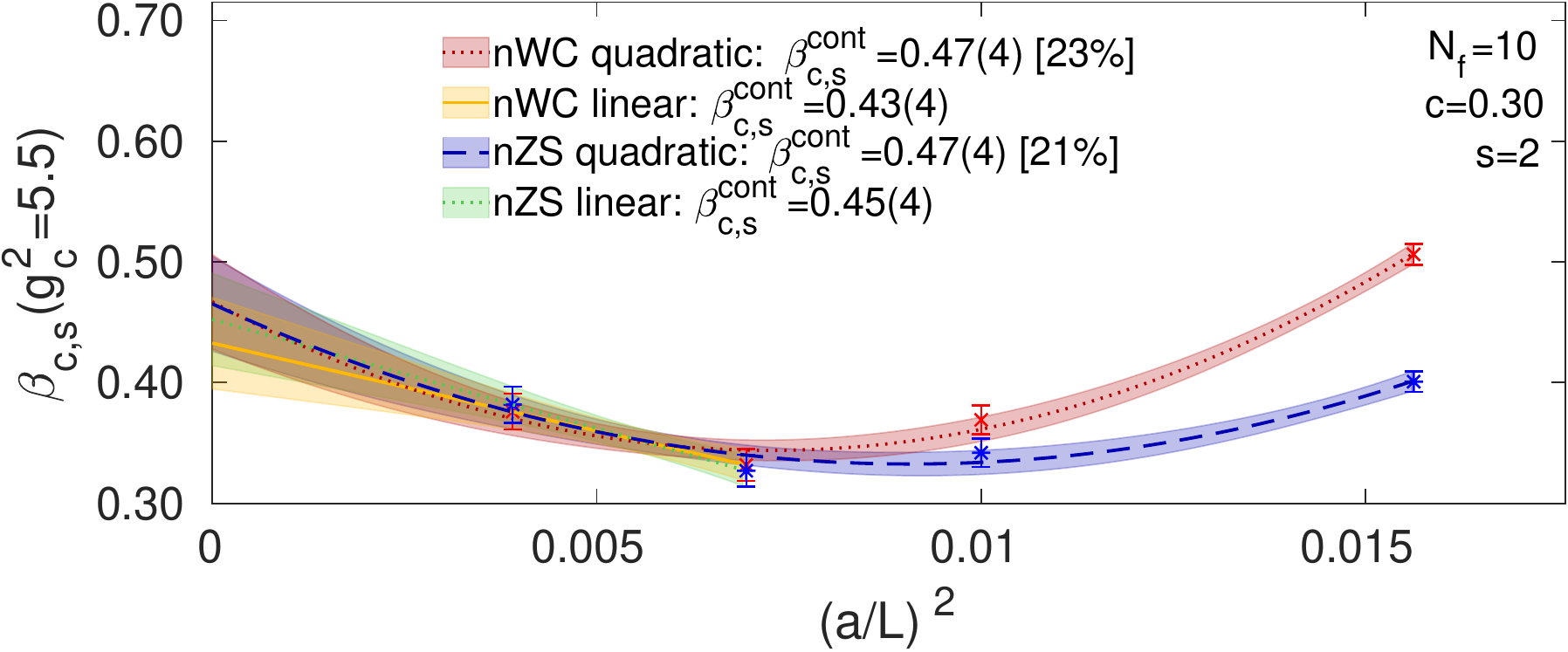}
  \caption{Ten flavor finite volume step-scaling functions for a scale change of $s=2$ at the renormalization scheme $c=0.3$. Colored symbols and bands correspond to individual lattice volume pairs; the black line with gray band shows  the continuum-limit discrete $\beta$-function obtained from a linear extrapolation of the three largest lattice volume pairs. An alternative extrapolation using a quadratic ansatz to fit all four lattice volume pairs is indicated by the dash-dotted line. The top panels show nZS and nWC data, while details of the continuum extrapolations are shown in the lower panels for $g_c^2=1.5$, 3.5, 5.5 with $p$-values in percent given in brackets. (Error bars and bands reflect only statistical uncertainties.)}
  \label{fig:Nf10}
\end{figure}

We analyze the $N_f=10$ system following the steps  detailed for twelve flavors. Our goal here is to verify the simulation and analysis by comparing to published DW results of \cite{Chiu:2016uui,Chiu:2017kza,Chiu:2018edw} in the weak coupling regime.  At present we have four lattice volume pairs, $8\to 16$, $10\to 20$, $12\to 24$ and $16\to 32$, with the largest lattice volume pair covering only a limited range, $g_c^2 \lesssim 6.0$. Motivated by the choice of Refs.~\cite{Chiu:2016uui,Chiu:2017kza,Chiu:2018edw}, we chose $c=0.3$ in this analysis and compare our preferred nZS combination to tree-level improved Wilson flow with clover operator (nWC). The outcomes of our analysis are the four sets of colored data points in the top two panels of Fig.~\ref{fig:Nf10}.
 Again, we first interpolate the individual lattice volume-pairs with a third order polynomial, then extrapolate to the $(a/L)^2\to 0$ continuum limit which we restrict to the range of our largest $16\to 32$ data set. We extrapolate to the continuum using a linear ansatz for the two largest lattice volume pairs (black solid line with gray error band) or use a quadratic ansatz to extrapolate all four lattice volume pairs (gray dash-dotted line)\footnote{Both nZS and nWC data can be perfectly well extrapolated for $g_c^2<4.0$ by fitting the two largest lattice volumes using only a constant term. At stronger couplings a slope in $(a/L)^2$ is required and well resolved. Hence we do not show the constant fit for $N_f=10$.}.

For selected values of $g_c^2$, we present details on the continuum extrapolation in the lower panels of Fig.~\ref{fig:Nf10}  showing in total four different extrapolations: linear and quadratic for our preferred nZS data, and for the nWC data favored in Refs.~\cite{Chiu:2016uui,Chiu:2017kza,Chiu:2018edw}. The simulations differ however in their choice of gauge action, we choose Symanzik whereas \cite{Chiu:2016uui,Chiu:2017kza,Chiu:2018edw} uses Wilson gauge action. Again we observe that  nZS data exhibit small discretization effects but nWC  is quite similar, as predicted by the small perturbative $O((a/L)^2)$ corrections.  All extrapolations agree within statistical uncertainties in the continuum.

\subsection{\label{sec:Discussion} Discussion}

\begin{figure*}[tb] 
  \centering
  \includegraphics[width=0.95\columnwidth]{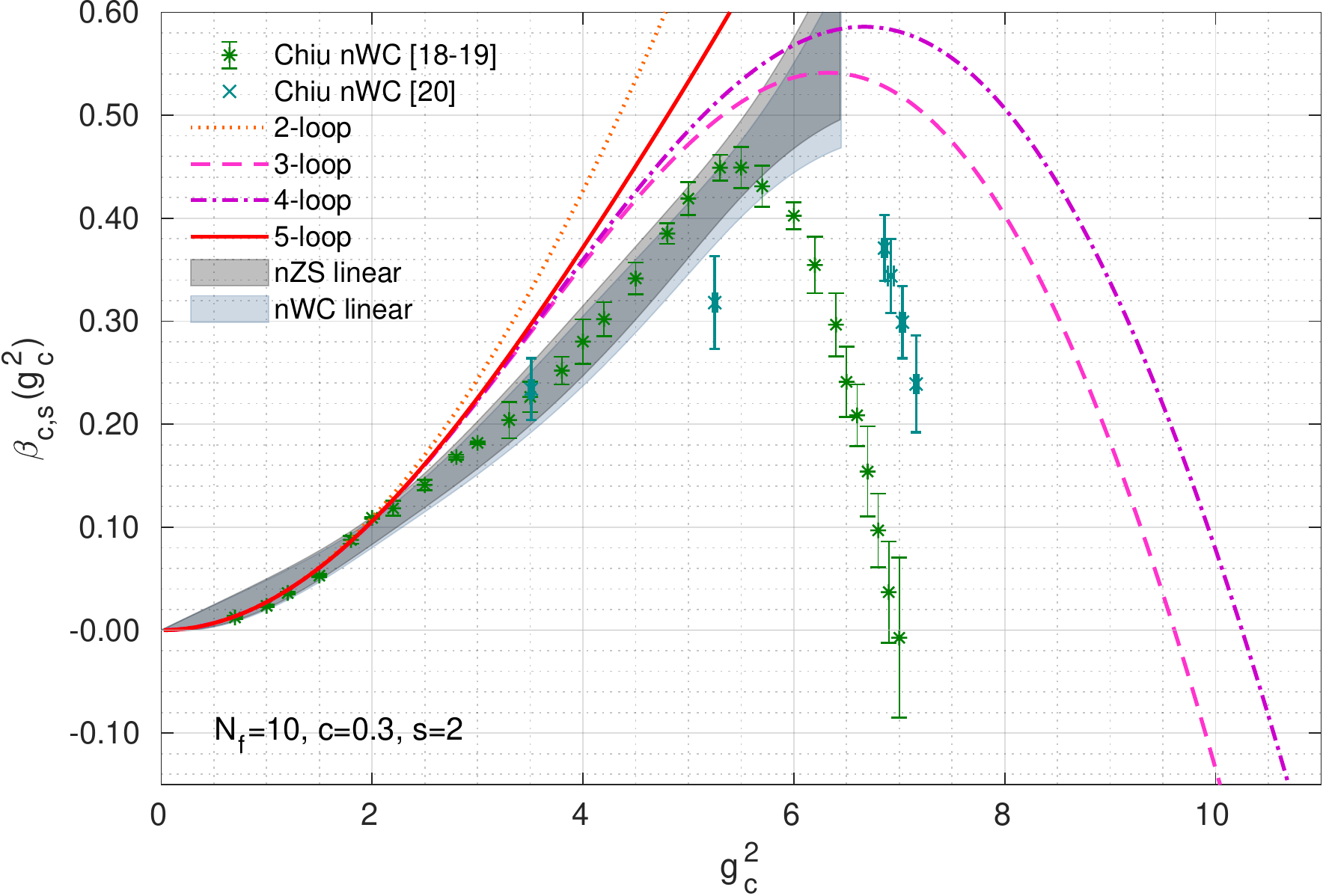}\hfill
   \includegraphics[width=0.95\columnwidth]{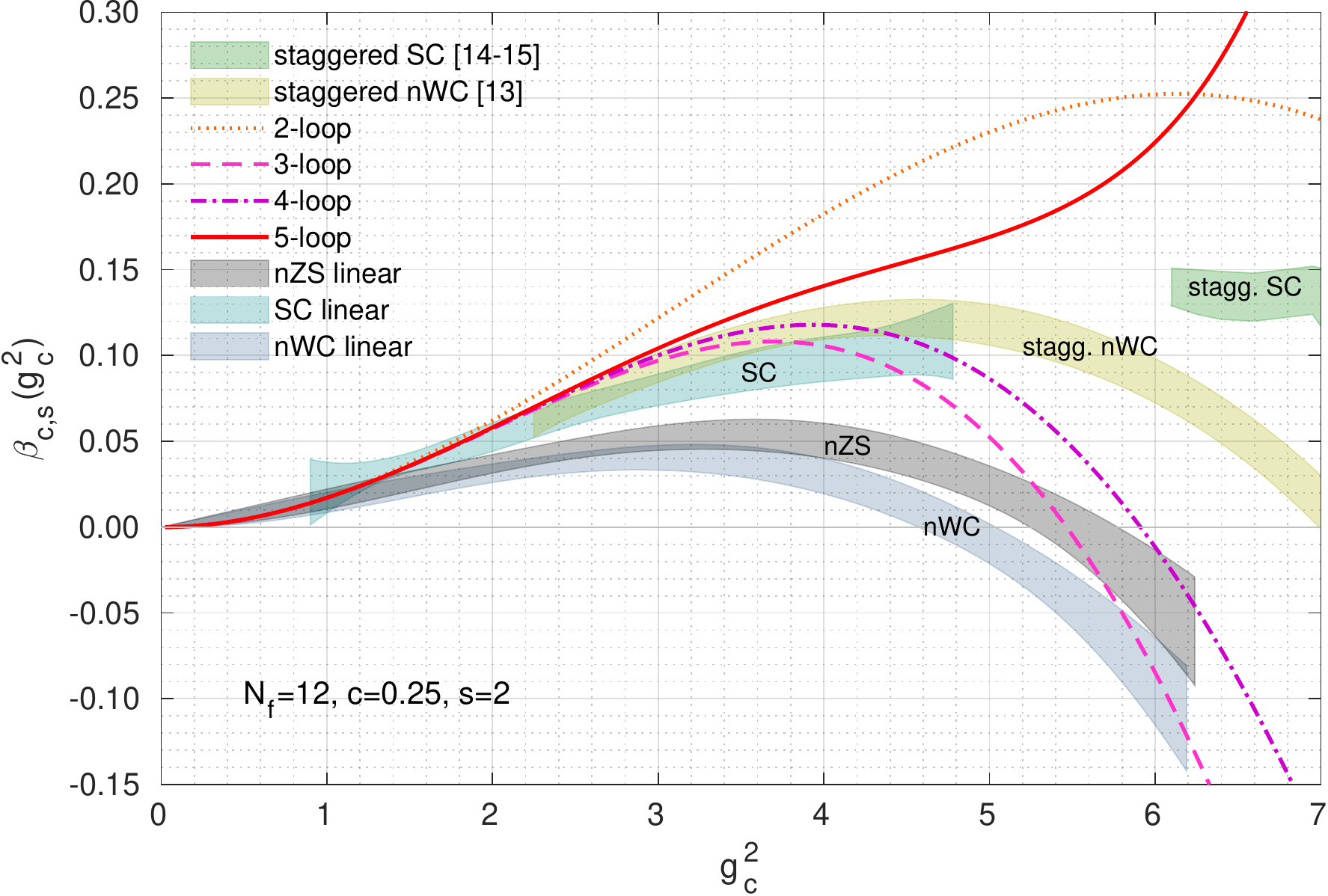}      
  \caption{Comparison of our continuum limit results (with statistical uncertainties only) for $N_f=10$ (left) and $N_f=12$ (right) to other non-perturbative \cite{Chiu:2016uui,Chiu:2017kza,Chiu:2018edw,Hasenfratz:2016dou,Fodor:2016zil,Fodor:2017gtj} and perturbative predictions \cite{Baikov:2016tgj,Ryttov:2016ner}. In addition to our preferred nZS determination, we show determinations using nWC and, in case of $N_f=12$, also SC. The latter has large discretization errors rendering its continuum limit extrapolations unreliable.}
  \label{fig:comp}
\end{figure*}

As demonstrated above, our preferred nZS data exhibit small cut-off effects and the continuum-limit results are consistent between different continuum  extrapolations and also agree with determinations based e.g.~on WW, nWC, nSC, which all have  relatively small discretization effects \cite{Nf12stepScaling}.  In Fig.~\ref{fig:comp} we show our preferred nZS continuum limit results  (gray error band) in comparison to other non-perturbative determinations  as well as to perturbative $\overline{\textrm{MS}}$  2, 3, 4, or 5-loop predictions~\cite{Baikov:2016tgj,Ryttov:2016ner}. For both $N_f=10$ and $N_f=12$ we observe that our DW result lies mostly below the 3-loop  prediction (pink dashed line).

Focusing on the results for $N_f=10$ shown in the plot on the left, we compare our findings for nZS and nWC to the nonperturbative determination by Chiu using optimal domain-wall fermions \cite{Chiu:2016uui,Chiu:2017kza,Chiu:2018edw}.  Our  nZS and nWC  continuum limit predictions closely overlap and for  $g_c^2 \lesssim 5.7$  are in excellent agreement with Chiu's 2016 result and within 2$\sigma$ of the  $g_c^2=5.25$ data point of his 2018 analysis. We refrain from speculating about the trend at stronger couplings because our data on the largest lattice volume-pair reach only to about $g_c^2\sim 6$. Hence we cannot comment on any downward trend and the possibility of an IRFP suggested by Chiu's results nor on the steady increase observed by the LatHC collaboration \cite{Fodor:2017gtj,Fodor:2018tdg}.  We are investigating the strong coupling region and will report our findings in a future publication. Nevertheless observing perfect agreement of two entirely independent determinations of $\beta_{c,s}$ over a wide range in $g_c^2$ boosts  confidence in our calculation.

\begin{figure}[tbp] 
  \centering
  \includegraphics[width=0.95\columnwidth]{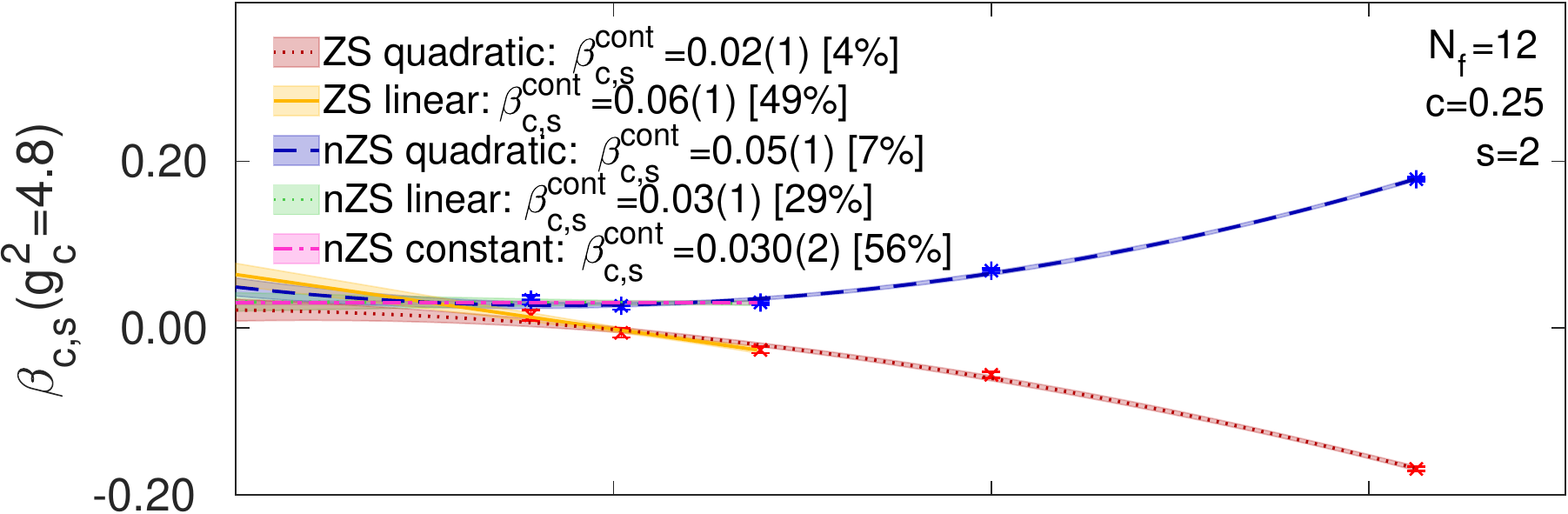}\\
  \includegraphics[width=0.95\columnwidth]{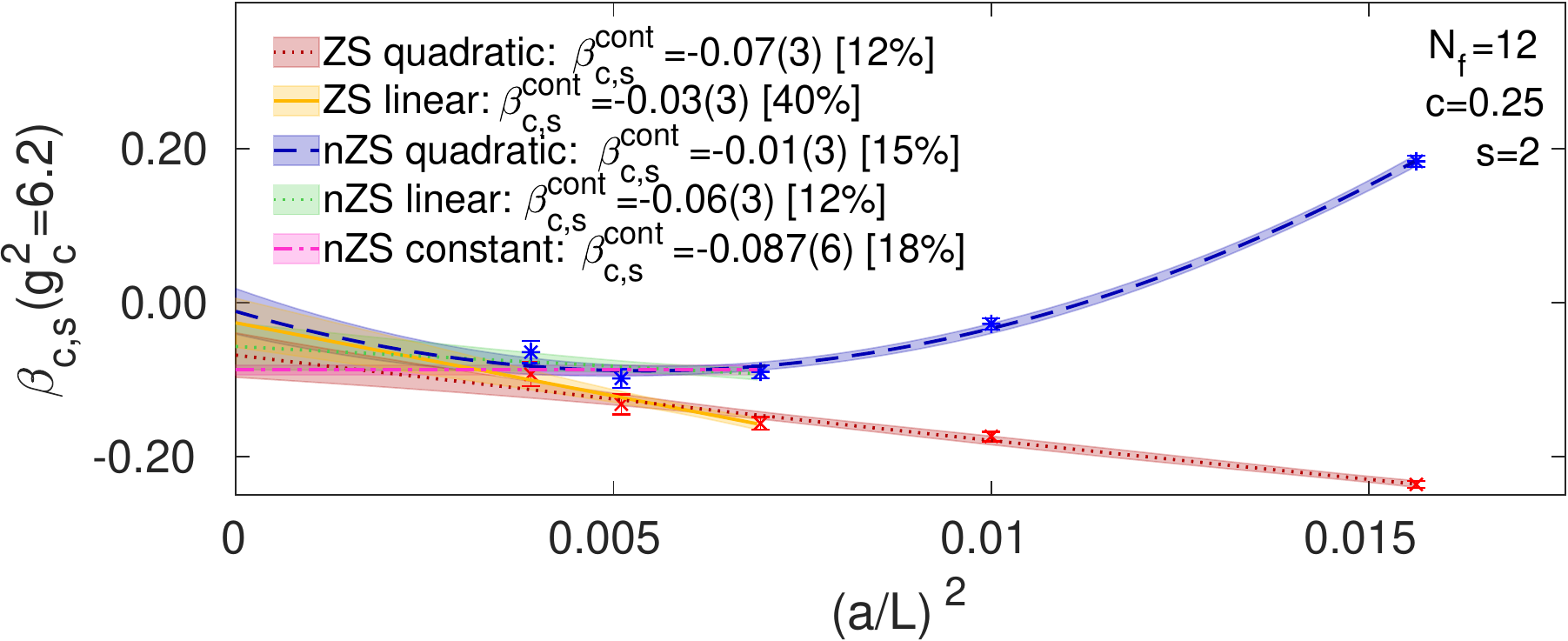}
  \caption{Continuum extrapolations of $N_f=12$ ZS data with (blue stars) and without (red crosses) tree-level normalization before/after the fixed point. (Error bars and bands reflect only statistical uncertainties.)}
  \label{fig:FP12}
\end{figure}

Results for twelve flavors are shown in the right panel of Fig.~\ref{fig:comp}. In addition to the perturbative $\overline{\textrm{MS}}$ predictions we also show the results obtained with staggered fermions from Ref.~\cite{Hasenfratz:2016dou} and Refs.~\cite{Fodor:2016zil,Fodor:2017gtj}. While at weaker couplings the staggered step scaling function is consistent among staggered calculations \cite{Cheng:2014jba,Hasenfratz:2016dou,Lin:2015zpa,Fodor:2016zil}, there is discrepancy between  Refs.~\cite{Fodor:2017gtj,Fodor:2016zil} and \cite{Hasenfratz:2016dou} at strong couplings, $g_c^2 \gtrsim 6.5$. References \cite{Fodor:2016zil,Fodor:2017gtj} predict a flat $\beta$ function with value around 0.14, whereas Ref.~\cite{Hasenfratz:2016dou} observes an IRFP around $g_c^2\sim 7.3$. Our findings based on the preferred nZS analysis identify an IRFP around $g_c^2\sim 5.5$ and seems to be in tension with both staggered predictions. In Fig.~\ref{fig:FP12}, we demonstrate that the step scaling function changes its sign by comparing continuum extrapolations before/after the fixed point using both the nZS and ZS analysis.  

In addition to the continuum limit obtained from our preferred nZS analysis, we show in Fig.~\ref{fig:comp} the continuum limit of the alternative SC and nWC analysis to match the flow/operator choices used in the staggered calculations. As mentioned before, the SC data exhibit large discretization effects, cover only a smaller range in $g_c^2<5.0$, and the extrapolation is less reliable because linear and quadratic fits differ by several sigma. Incidentally, the SC data extrapolate to a $\beta$ function which is close to the staggered prediction of Ref.~\cite{Hasenfratz:2016dou} and similar in magnitude to the result of Refs.~\cite{Fodor:2016zil,Fodor:2017gtj} obtained for $g_c^2>6.0$. While References \cite{Fodor:2016zil,Fodor:2017gtj} also use SC with Symanzik gauge action, the determination of Hasenfratz and Schaich \cite{Hasenfratz:2016dou} is based on nWC with Wilson gauge gauge action. Considering the nWC data with relatively small discretization effects, we find a continuum limit prediction which is consistent with our preferred nZS analysis at the $\sim 1\sigma$ level.

By comparing our three different continuum limit predictions, the right plot in Fig.~\ref{fig:comp} demonstrates how significant discretization effects can be and that using a fully $O(a^2)$ improved set-up is advantageous. Note however that so far no staggered determination uses a fully $O(a^2)$ improved set-up.  Additional investigations including estimates for systematic effects due to flow/operator choices as well as the dependence on the scheme $c$ are required to resolve the discrepancy between the present results. 

To gain further insight into this discrepancy, we point out that discretization effects in data from DW simulations can be successfully reduced by taking advantage of the perturbative tree-level normalization.  For simulations with staggered fermions and Symanzik gauge action the perturbatively preferred combinations are nSS or nZS \cite{Fodor:2014cpa}.  Reference \cite{Fodor:2017gtj}, however, finds that the perturbatively poor SC combination shows smaller cut-off effects.  A priori there is no expectation for a perturbative improvement to work in a nonperturbative system; observing its success for one type of fermions and its failure for a different type may signal that one of the simulations is (much) closer to the perturbative FP than the other. It is further interesting  to consider the expectation value of the plaquette. The plaquette, the smallest Wilson loop, is a good indicator of UV fluctuations. For the range of bare couplings covered in our DW simulations, we find  an average plaquette (normalized to 1) in the range $0.6-0.8$, in contrast to  the range $0.3-0.6$ of the staggered simulations of  Ref.~\cite{Hasenfratz:2016dou}.  A QCD simulation with an average plaquette of 0.3 would lie outside the scaling window of the Gaussian FP and continuum extrapolations would at least be difficult.

If even after resolving systematic effects due to different flow/operator choices predictions based on DW and staggered fermions disagree, this might suggest that either they are not governed by the same fixed point or that one or both of the simulations are not in the basin of attraction of the perturbative Gaussian and/or the emerging infrared fixed point. The continuum limit extrapolation should force the simulations close to the Gaussian FP, but due to the slowly running gauge coupling of the  $N_f=12$  system that might not be the case in practice.

The DW data for $N_f=12$ deviate from the perturbative 2-loop prediction for $g_c^2 > 1.5$ and stays significantly below the perturbative predictions. It is worth pointing out that the finite volume step scaling function is only 1-loop universal and $\be_{c,s}$ also depends on the scheme $c$ and the scale change $s$.  Thus deviations from the perturbative  $\overline{\textrm{MS}}$ curves have no significance to the reliability of the numerical results.

\section{Conclusion}\label{sec:conclusion}
Universality of DW,  Wilson, and staggered fermions at the Gaussian $g^2=0$ fixed point is subtle, but can be proven perturbatively.  Universality  of a conformal  system requires that taste symmetry of staggered fermions stays restored  even at finite gauge coupling along the renormalized trajectory connecting the Gaussian and conformal FPs, and that the simulations are done in the vicinity (basin of attraction) of this renormalized trajectory.

Motivated by these observations and the importance of conformal and  near-conformal models for  BSM phenomenology, we have  studied the gradient flow step scaling function of $N_f=10$ and 12 fundamental flavors with  domain wall fermions. The gradient flow step scaling function depends on the renormalization condition but should not depend on the regularization.  Our analysis  features the fully ${\cal O}(a^2)$ improved set-up based on Symanzik gauge action, Zeuthen flow, and Symanzik operator, combined with tree-level normalization. Indeed  we find that this combination has very small cut-off effects.  Our ten flavor results are consistent with the independent  DW calculation by Chiu~\cite{Chiu:2016uui,Chiu:2017kza,Chiu:2018edw} for $g^2_c\lesssim 5.7$, validating our approach.  In case of $N_f=12$, our preferred nZS analysis differs from staggered predictions. Taking advantage of alternative determinations based on different flow/operator combinations (SC and nWC), we demonstrate that in particular SC exhibits large discretization effects, whereas nWC agrees with nZS at roughly the $1\sigma$ level. To understand the difference between results based on domain wall and staggered fermions, it is hence important to estimate systematic effects due to flow/operator choices. We emphasize that the ``true'' continuum limit should be independent of the choices for flow, operator, or action.

If significant differences persist even after properly accounting for all systematic effects, this may imply  that at least one of the simulations is not in the basin of attraction of the fixed point.  DW simulations exhibit the correct flavor symmetry, have smaller cut-off effects/UV fluctuations, and feature successful perturbative improvement \`a la Symanzik, increasing our confidence in the result presented here. 

Our simulation of the $N_f=12$ system predicts an infrared fixed point around $g^2_c \sim 5.5$ in the $c=0.25$ scheme where the step scaling function changes sign. We do not have sufficient strong gauge coupling data in the $N_f=10$ system  to comment on the possible existence of an IRFP.

\section*{Acknowledgments}
We are very grateful to Peter Boyle, Guido Cossu, Antonin Portelli, and Azusa Yamaguchi who developed the \texttt{Grid} software library providing the basis of this work and who assisted us in installing and running \texttt{Grid} on different architectures and computing centers. In addition we would like to thank Alberto Ramos and Stefan Sint for help in understanding the Zeuthen flow and Andrew Pochinsky for support in implementing Zeuthen flow in \texttt{qlua}. We are grateful to Ting-Wai Chiu for sharing his results on the $N_f=10$ $\beta$ function enabling us to show direct a comparison with our findings. Computations for this work were carried out in part on facilities of the USQCD Collaboration, which are funded by the Office of Science of the U.S.~Department of Energy and the RMACC Summit supercomputer, which is supported by the National Science Foundation (awards ACI-1532235 and ACI-1532236), the University of Colorado Boulder, and Colorado State University, and on \texttt{stampede2} at TACC allocated under the NSF Xsede program to the project TG-PHY180005. We thank Brookhaven National Laboratory, Fermilab,  Jefferson Lab, the University of Colorado Boulder, TACC, the NSF, and the U.S.~DOE for providing the facilities essential for the completion of this work.  A.H.  and O.W. acknowledge support by DOE grant DE-SC0010005 and C.R. by DOE grant DE-SC0015845.   This project has received funding from the European Union's Horizon 2020 research and innovation programme under the Marie Sk{\l}odowska-Curie grant agreement No 659322. 
\appendix

\section{Residual chiral symmetry breaking}

\begin{figure}[tb]
    \centering
  \includegraphics[width=0.95\columnwidth]{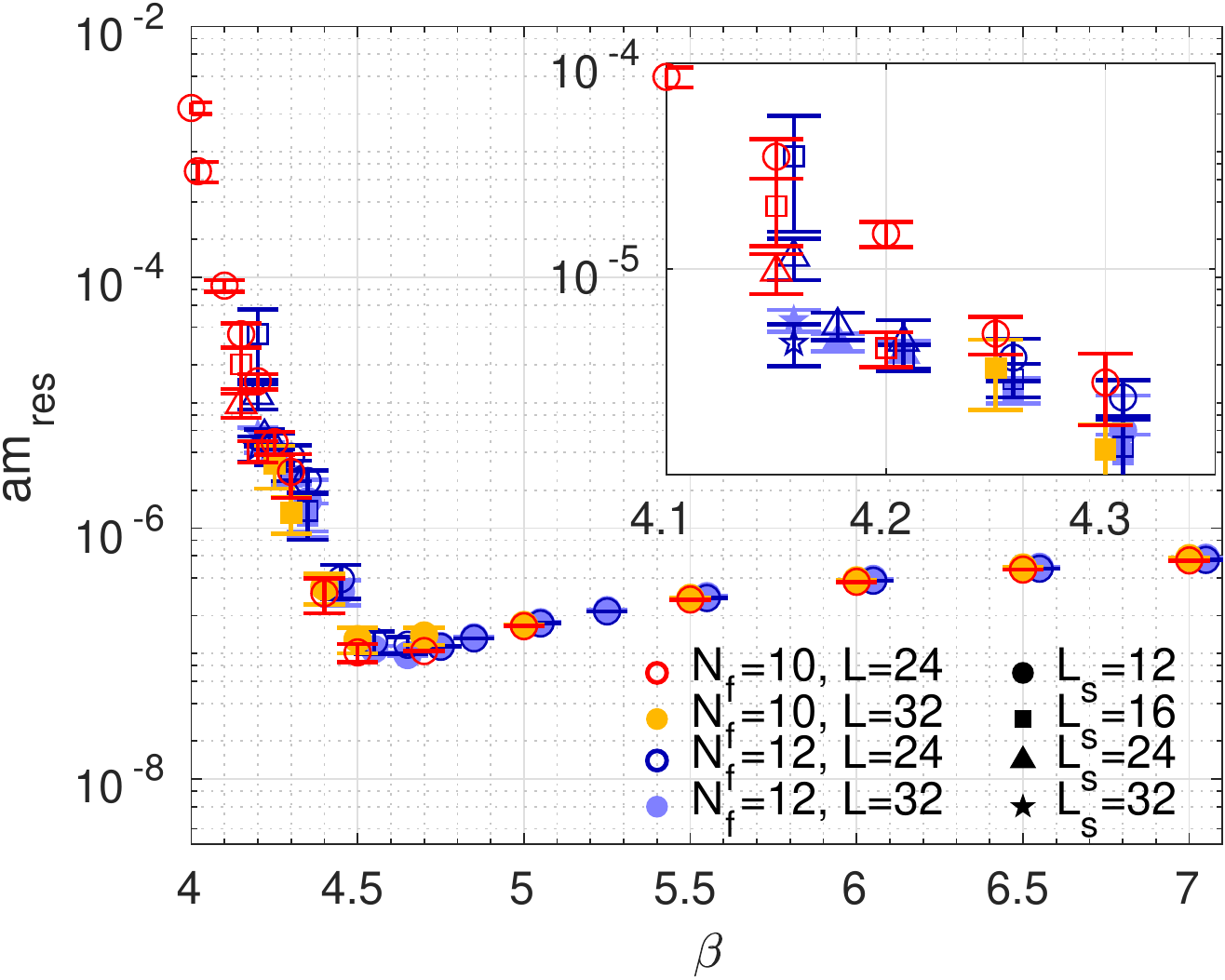}  
  \caption{Measured residual chiral symmetry breaking parametrized by $a m_\text{res}$ for simulations with $N_f=10$ and $12$ flavors on $L/a=24$ and 32 volumes with $L_s=12$, 16, 24, and 32. Data for $N_f=12$ are shown with a small horizontal offset to the right for better visibility and the range $4.1\le \beta \le 4.35$ is enlarged in the top right corner.  Only statistical uncertainties are shown.}
\label{Fig.mres}
\end{figure}

The residual chiral symmetry breaking is  parametrized by $m_\text{res}$  which we  estimate using the ratio of the midpoint-pseudoscalar over the pseudoscalar-pseudoscalar correlator. In Fig.~\ref{Fig.mres} we show $a m_\text{res}$ as the function of the bare gauge coupling $\be$ both for  ten and twelve flavors on lattice volumes with $L/a=24$ and 32.  Most ensembles are generated with $L_s=12$, shown by the circles in the plot. First of all, we observe that $a m_\text{res}$ grows rapidly for stronger couplings/decreasing values of $\beta$ and within our (statistical) uncertainties $a m_\text{res}$ is independent of the lattice volume. Secondly and more interesting, we observe that $a m_\text{res}$ does not  exhibit a noticeable dependence on the number of flavors used to generate the dynamical gauge field ensembles. 

\label{Appendix.mres}

As can be seen in Fig.~\ref{Fig.mres}, for $\beta \ge 4.4$ the residual mass is smaller than $10^{-6}$ i.e.~it is zero within the limits of single precision. For simulations at stronger coupling, we increase $L_s$ from 12 up to 32 at $\beta=4.15$ and maintain a residual mass of $5\cdot 10^{-6}$.

Furthermore, we explicitly check for effects due to a finite residual mass by generating selected ensembles using different values of $L_s$. At our strongest coupling ($\beta=4.15$) for $N_f=12$ simulations, we generated e.g.~$L/a=8$ ensembles with $L_s= 12,\, 16,\, 32$ and $L/a=16$ ensembles with $L_s= 12,\, 24,\, 32$. Using nZS data we obtain $\beta_{c,s}$ values at $c=0.25$ for all nine combinations for the lattice volume pair $8\to 16$ in Eq.~(\ref{Eq.mres8_16}) and use the $L/a=16$ ensembles together with $L/a=32,\,L_s=32$ in Eq.~(\ref{Eq.mres16_32}) to determine $\beta_{c,s}$ for $16\to 32$.
\begin{align}
  &\begin{matrix}
    &   & \multicolumn{3}{c}{L/a=8}\\
    &L_s   & 12        & 16        & 32 \\  
    \multirow{3}{*}{\rotatebox{90}{$L/a=16$}}
    &12 & \phantom{+}0.184(16) & \phantom{+}0.176(12) & \phantom{+}0.174(12) \\
    &24 & \phantom{+}0.222(16) & \phantom{+}0.214(11) & \phantom{+}0.212(12) \\
    &32 & \phantom{+}0.230(17) & \phantom{+}0.222(12) & \phantom{+}0.220(13) \\
  \end{matrix}  
  \label{Eq.mres8_16}\\[3mm]
  &\begin{matrix}
    \multirow{3}{*}{\rotatebox{90}{$L/a=32$}}     
    &   & \multicolumn{3}{c}{L/a=16}\\
    &L_s   & 12        & 24        & 32 \\
    & 32 & -0.070(33)  & -0.108(33) & -0.116(33)\\
   \end{matrix}
    \label{Eq.mres16_32}
\end{align}

Equation (\ref{Eq.mres8_16}) demonstrates that our preferred choice of $L_s=16$ for $L/a=8$ and $L_s=24$ for $L/a=16$ is sufficient at $\beta=4.15$ because $\beta_{c,s}$ values are consistent with values corresponding to ensembles with larger $L_s$. Likewise we cannot resolve a difference for $L/a=16$ with $L_s=24$ or $32$ for the $16\to 32$ pair in Eq.~(\ref{Eq.mres16_32}) and take that as guidance to keep $am_\text{res} \lesssim 10^{-5}$. This also ensures the residual mass is much smaller than the energy scale introduced by the volume, $m_\text{res} \ll  L^{-1}$, i.e.~the configurations are guaranteed to be volume squeezed instead of finite-mass deformed.

\newpage
\section{\texorpdfstring{nWC $N_f=12$ step-scaling function}{nWC Nf=12 step-scaling function}}
\label{Sec.Nf12nWC}

\begin{figure}[h!] 
  \centering
  \includegraphics[width=0.95\columnwidth]{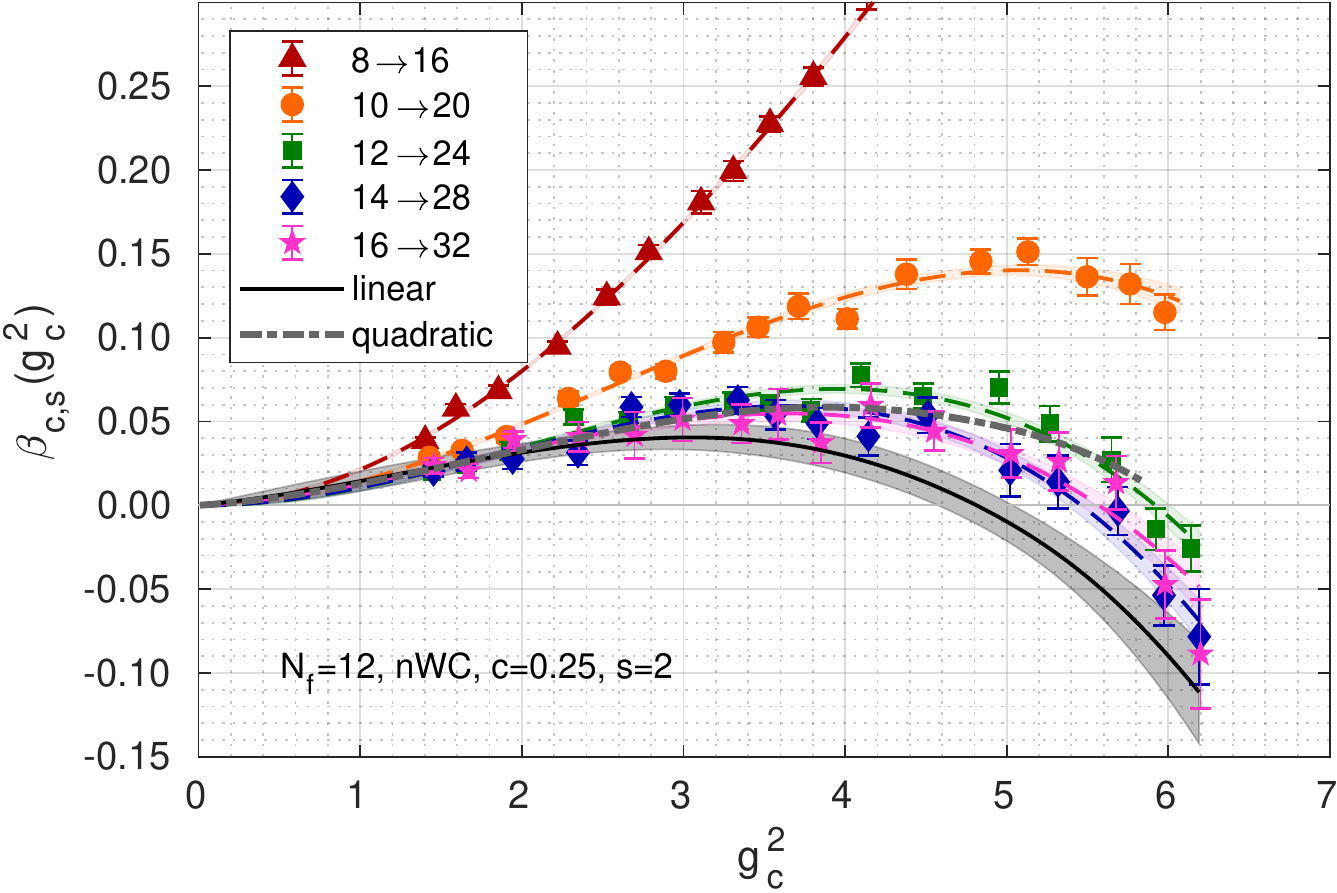}\\
  \includegraphics[width=0.95\columnwidth]{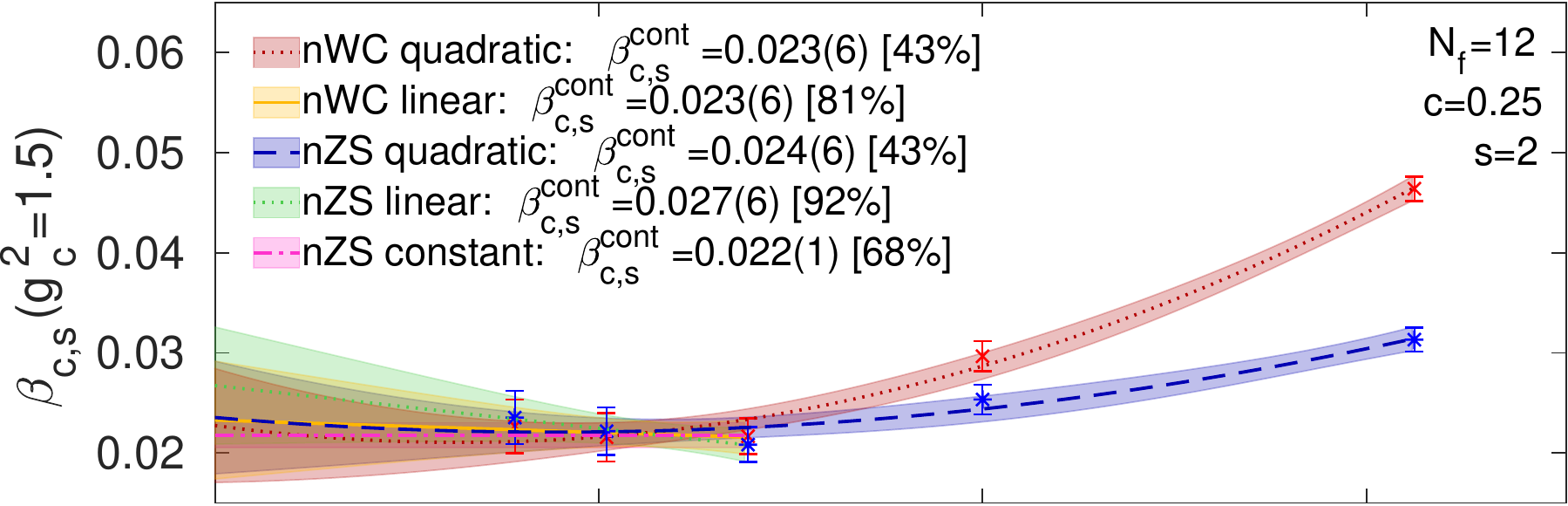}\\
  \includegraphics[width=0.95\columnwidth]{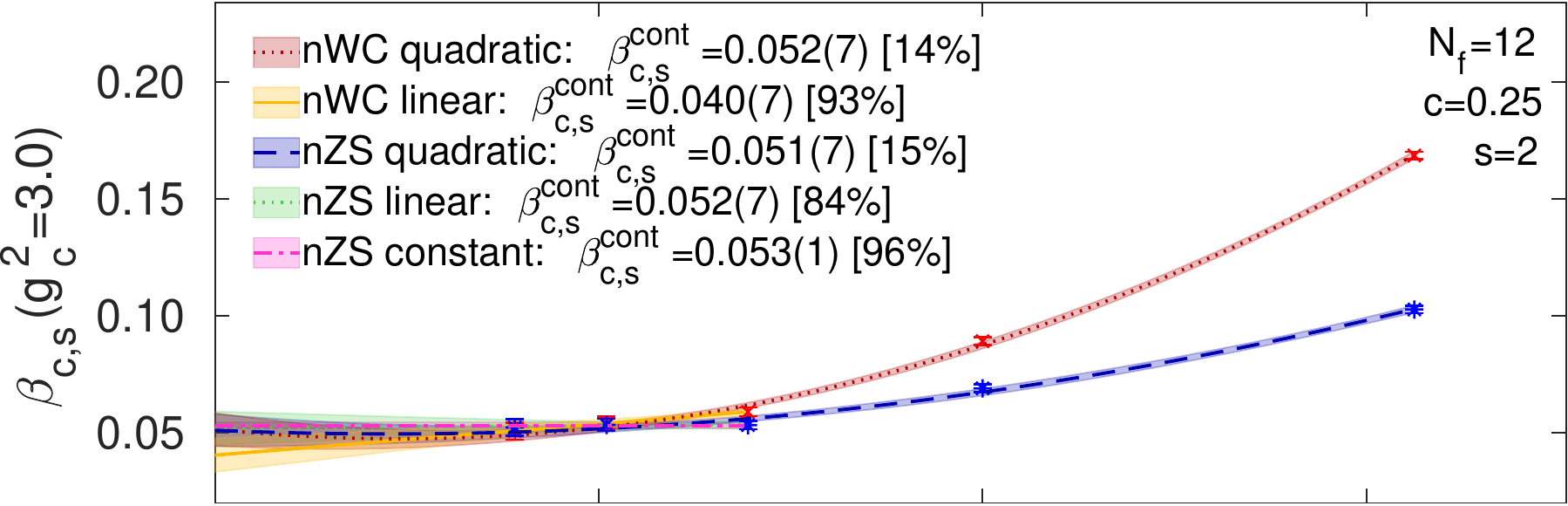}\\
  \includegraphics[width=0.95\columnwidth]{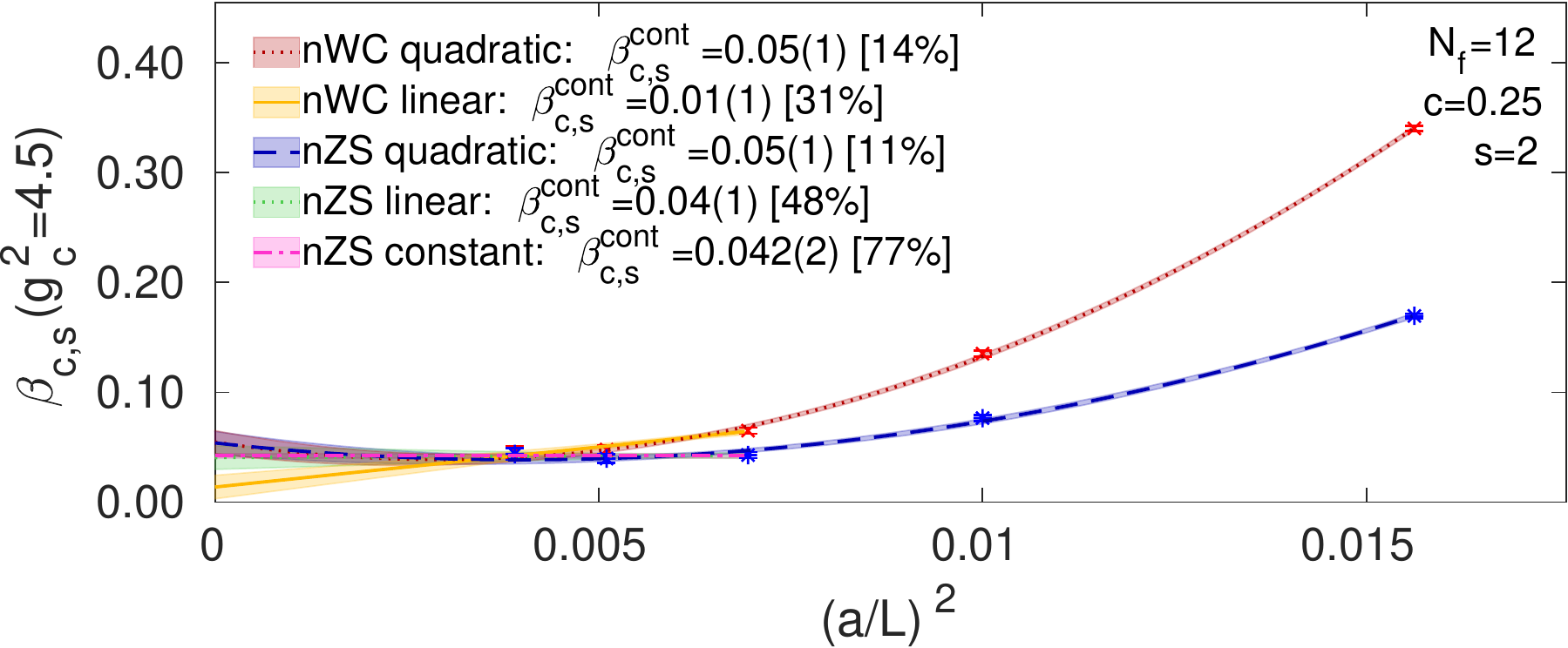}
  \caption{Step scaling functions for degenerate 12 flavors analyzed  in the renormalization scheme $c=0.25$ using the nWC combination. Conventions are the same as in Figs.~\ref{fig:Nf12} and \ref{fig:Nf10}.}
  \label{fig:Nf12_nWC}
\end{figure}

\section*{References}
\bibliography{../General/BSM}
\bibliographystyle{elsarticle-num}

\end{document}